\documentclass[10pt, conference]{IEEEtran}
\IEEEoverridecommandlockouts

\usepackage{cite}
\usepackage{amsmath,amssymb,amsfonts}
\usepackage{algorithmic}
\usepackage{graphicx}
\usepackage{textcomp}
\usepackage{xcolor}
\usepackage[breakable]{tcolorbox}
\usepackage{enumitem}
\usepackage{float}

\tcbset{
    colback=white,
    colframe=black!50,
    left=3pt,
    right=3pt,
    top=3pt,
    bottom=3pt,
    boxsep=2pt,
    before=\vspace{3pt},
    after=\vspace{3pt},
    width=\columnwidth,
    title=26pt,
    fontupper=\small,
    fontlower=\small
}
\def\BibTeX{{\rm B\kern-.05em{\sc i\kern-.025em b}\kern-.08em
    T\kern-.1667em\loIr.7ex\hbox{E}\kern-.125emX}}
\begin{document}

\title{A Human-Centric Requirements Engineering Framework for Assessing Github Copilot Output
}

\author{\IEEEauthorblockN{Soroush Heydari}
\IEEEauthorblockA{\textit{Ontario Tech University}\\
Oshawa, Canada \\
soroush.heydari@ontariotechu.net}

}

\maketitle

\begin{abstract}
The rapid adoption of Artificial Intelligence(AI) programming assistants such as GitHub Copilot introduces new challenges in how these software tools address human needs. Many existing evaluation frameworks address technical aspects such as code correctness and efficiency, but often overlook crucial human factors that affect the successful integration of AI assistants in software development workflows. In this study, I analyzed GitHub Copilot's interaction with users through its chat interface, measured Copilot's ability to adapt explanations and code generation to user expertise levels, and assessed its effectiveness in facilitating collaborative programming experiences. I established a human-centered requirements framework with clear metrics to evaluate these qualities in GitHub Copilot chat. I discussed the test results and their implications for future analysis of human requirements in automated programming. \footnote{This paper was written as a final project deliverable for the graduate software engineering course Foundations of Software Engineering, taught at Ontario Tech University by Dr. Sanaa Alwidian in Fall 2024. The paper was completed and submitted in November 2024.}
\end{abstract}


\section{Introduction}

\subsection{Problem}

Traditional software quality metrics fail to cover essential human aspects such as code comprehensibility, collaborative capabilities, and inclusivity, factors that directly influence a development tool's effectiveness in real-world scenarios \cite{b3}. There are few standardized frameworks for evaluating how well AI programming assistants meet human-centric requirements, restricting the testing and improvement of their ability to serve diverse user needs. This creates a significant challenge for both developers and organizations implementing these tools, as they cannot effectively evaluate whether an AI assistant will fully meet user requirements \cite{b4}.

\subsection{Objectives}

The objective was to propose and validate a requirements engineering framework that measured GitHub Copilot's ability to integrate human aspects into its responses. This framework evaluated human-centered software qualities such as inclusivity and comprehensibility within Copilot chat's output. The evaluation contributed to a broader goal of creating more inclusive, understandable, and effective AI programming assistants that consider the human needs of the software developers that use them.
\\
GitHub Copilot was chosen for this study because it is a Large Language Model (LLM) specifically geared towards code generation and project collaboration. It is also widely used and marketed to students. As such, Copilot must pass human requirements to be an effective tool for use by students and development teams.
\\
User stories were included in the testing process to represent the stakeholder profile involved with Copilot. User stories are a tool used to describe software system interactions from the perspective of one or more stakeholders. A user story consists of an actor whose profile and needs are written in the story, an action that represents the tasks the users will do, and a goal that describes what the user wishes to gain from the action \cite{b13}. A user story is often combined with a set of acceptance criteria, which are used to determine whether the user's stated goal and needs were fulfilled. User stories are widely used in Software Requirements Engineering (SRE), particularly in Agile development. A user is structured as a set of phrases: "As a [user type], I want [objective] so that [benefit]". User Stories have been used to measure human-centric qualities such as fairness \cite{b13}. They provide a degree of personalization that allows for context-aware testing of non-functional and human requirements. 

The user story in this study is meant to represent a profile of a novice programmer using GitHub Copilot chat as a pair programming aid for a coding assignment. The user story's details influenced the content of the prompts used in this study. These details are explained further in the Methodology section. 

\section{Related Work}

Human aspects refer to a set of characteristics such as personality, motivation, emotions, communication, collaboration, and culture that are present in the software development cycle, and that stakeholders such as software practitioners and end-users express a need for \cite{b3}\cite{b5}\cite{b6}. Personality in a software engineering team correlates with positive Requirements Engineering (RE) outcomes, and demographic diversity is an ongoing challenge in software engineering \cite{b5}\cite{b6}. Insufficient communication can compromise requirement accuracy, lack of collaboration can obstruct implementation, and lack of motivation reduces the quality of deliverables. Other important qualities include deliverable trustworthiness, code clarity, accessibility and inclusion for both users and developers. Failure to consider these can negatively impact software development, maintenance, and user experience. In short, these aspects are crucial parts of the software development cycle.

Frameworks such as Grounded Theory have been developed to study human
aspects and their links to functional requirements in software engineering \cite{b7}. Grounded Theory provides clear qualitative metrics and an integrated approach for measuring human aspects in software development. This could help bridge existing research gaps relating to human-centric requirements engineering, such as complexity of needs, contextual changes, and the subjective nature of measuring human aspects \cite{b3}. Frameworks such as Grounded Theory are crucial for interpretation as human aspects cannot be measured in isolation, and software development exists within the context of broader society. 

Evaluation standards for automated programming output commonly include static analyzers such as Pylint and Checkstyle that check code type and style to flag potential runtime issues \cite{b2}\cite{b8}. Some metrics also deal with indirect aspects such as poor design and incomprehensibility (referred to as code smells), with the most common example being multiple-nested containers. Tools such as Pysmell are often used to detect and resolve these quality issues \cite{b1}. These static analyzers have clear limits; one study found that 53\% of passing Java code snippets and 37\% of passing Python code snippets demonstrated maintainability issues later on, proving that functional test cases are insufficient for quality testing automated programs \cite{b9}. Existing technical quality methods are capable of resolving syntax and runtime errors, but incapable of guaranteeing high-quality, usable, and maintainable code. This suggests that a more comprehensive software quality measurement framework is needed, one that relies on more than just functional tests. 

These studies stated that human evaluation is required to assist interpretation in order to avoid false positives which can occur at up to a rate of 25\% \cite{b8}\cite{b10}. The assistant-like role of automated programming tools and their reliance on human partners further shows the necessity of including human aspects in the software quality review process. Any factor that could compromise the user or developer experience of the user or developer could also lower software quality.

Some studies discussed human aspects in a more specific context. One study explored the environmental costs of LLM training and output and discussed how these costs must be balanced against code quality \cite{b11}. The study concluded that LLMs do not outperform humans at writing more environmentally sustainable code when energy use and pollution are taken into account. Other studies evaluated sensitivity and cultural context in LLM output. They noted that LLMs struggle with cultural nuance when determining requirements, and often used a one-size-fits-all approach \cite{b4} \cite{b12}. Some frameworks measured LLM output qualities such as code readability \cite{b13} \cite{b14}. Evaluations of LLM capabilities in more advanced software development tasks showed that they struggle with code maintenance and project management and can only accomplish such tasks when prompted extensively \cite{b6}. These studies suggested that LLMs are most effective when used collaboratively. As such, quality analysis for LLM programming assistants requires measuring their collaborative ability, including human factors.

One truth these studies showed is that human oversight is required to overcome algorithmic limitations in automated programming. Each study fixates on one individual human aspect and its relation to automated programming, showing a lack of integration of human-centered evaluation processes. Where solutions are discussed and proposed, they often describe human aspects in vague categorical terms. Despite the prevalence of frameworks to measure individual human aspects in an automated programming system, there is still no consensus on how to apply human requirement engineering to automated programming in a systematic way. This provides space for further research into more comprehensive frameworks that can measure the quality of human aspects in automated programming and incorporate them into software quality assessments.

\section{Methodology}

\subsection{Scope}
This study's scope was limited to the measurement of human centric requirements. The evaluation checklist and test cases concerned criteria that are not easily quantifiable and lacked established measurement and integration standards. Time limitations also influenced the boundaries of the scope. The main purpose was to gather measures of disparate individual human requirements identified in the literature review into a comprehensive checklist and test them against generated code.

The methodology used the following user story as a model for a stakeholder with human-centric requirements. The resulting Copilot prompt scenario was designed with this user story in mind.

\begin{table}[H]
\caption{User Story and Acceptance Criteria}
\begin{tabular}{|p{0.95\columnwidth}|}
\hline
\textbf{User Story} \\
As a new software developer \\
I want GitHub Copilot to assist me \\
So that I may develop a Flask web application \\
\hline
\textbf{Acceptance Criteria} \\
Given: The user has a GitHub account and Copilot chat opened. \\
When: The user inputs their requests into Copilot chat \\
Then: Copilot should return an appropriate and helpful solution, along with an instructive explanation. \\
\hline
\end{tabular}
\end{table}

This description was chosen to represent a hypothetical GitHub Copilot user that could not solely rely on technical output due to their lack of domain knowledge. As a result, the user in this story relied on clear and effective communication from the LLM to carry out tasks.

\subsection{Evaluation Criteria}
The AI-generated code was evaluated across four key dimensions of human-centric requirements:

\subsubsection{Inclusivity}
\begin{itemize}
    \item Cultural neutrality in variable names and comments
    \item Suitability across different user expertise levels
    \item International language support for code explanations
\end{itemize}

\subsubsection{Comprehensibility}
\begin{itemize}
    \item Intuitive, meaningful, and consistent variable naming
    \item Logical and easily followable code structure
    \item Clear explanations of code
\end{itemize}

\subsubsection{Collaboration}
\begin{itemize}
    \item Relevance of AI output to user context and needs
    \item Responsiveness to user actions
    \item Reasoning provided by AI for its output
\end{itemize}

\subsubsection{Domain Knowledge Integration}
\begin{itemize}
    \item Appropriate use of domain-specific terminology
    \item Handling of domain-specific edge cases
    \item Facilitation of domain expert review
\end{itemize}

Inclusion of elements in the checklist was considered in terms of whether their absence would cause a demonstrable negative impact. The intention was that each topic in the checklist must be at least partially validated at minimum for the tested model to meet human requirements.

\subsection{Test Cases}
Each dimension was evaluated through specific test cases designed to  assess compliance with human-centric requirements. The test cases were observable (clearly determined), repeatable (could be consistently performed), comprehensive, and actionable (failed tests indicated specific improvements are needed).

\subsubsection{TC1: Inclusivity Assessment}
\textbf{Objective:} Verify Copilot output is user-inclusive\\
\textbf{Input:} Generated code and original requirements\\
\textbf{Test Steps:}
\begin{enumerate}
    \item Assess prompt output for culturally specific references
    \item Determine if explanations are expertise-neutral
    \item Test criteria across multiple languages
\end{enumerate}

\subsubsection{TC2: Comprehensibility Assessment}
\textbf{Objective:} Evaluate Copilot output comprehensibility\\
\textbf{Input:} Generated code and their explanations\\
\textbf{Test Steps:}
\begin{enumerate}
    \item Review naming conventions for consistency
    \item Analyze code structure for logical flow
    \item Assess explanation quality and utility
\end{enumerate}

\subsubsection{TC3: Collaboration Assessment}
\textbf{Objective:} Verify effective AI-user interaction\\
\textbf{Input:} Complete interaction session including prompts and responses\\
\textbf{Test Steps:}
\begin{enumerate}
    \item Evaluate requirement-solution alignment
    \item Assess responsiveness to user feedback
    \item Review quality of explanatory reasoning
\end{enumerate}

\subsubsection{TC4: Domain Knowledge Assessment}
\textbf{Objective:} Verify appropriate application of domain expertise\\
\textbf{Input:} Generated code, requirements, and domain documentation\\
\textbf{Test Steps:}
\begin{enumerate}
    \item Verify domain-specific terminology usage
    \item Check edge case handling
    \item Assess documentation quality for expert review
\end{enumerate}

\subsection{Evaluation Metrics}
All test cases undergo manual review, with each criterion scored on a ternary scale:
\begin{itemize}
    \item 1.0: Unambiguous compliance
    \item 0.5: Ambiguous compliance
    \item 0.0: Unambiguous non-compliance
\end{itemize}

A score of 7.0 out of 12.0 was considered an acceptable grade for human requirements, as this demanded at least two unambiguous passes and discouraged unambiguous failures. A benchmark of 50 percent (6.0 out of 12.0) would have allowed code with completely ambiguous evaluations to pass. Because a score of 0.5 in this framework meant any number between 0.01 and 0.99, this minimum final acceptance score reduced the negative impact of ambiguity on numerical evaluation. Requirement failures were documented and maintained to help improve the automated programming system. 

Test input consisted of prompts that included both programming code and natural language explanations as Copilot output. The prompts were in the format of "I wish to accomplish X, show me how to do so and explain along the way". The output was tested against the human requirements checklist, which contained criteria covering both programming code and accompanying code explanations, including both paragraph-form explanations and in-line code comments. 

\section{Results and Analysis}

\subsection{User Story Prompt Results}

The user story was tested in Copilot's chat feature as follows. \\

\begin{tcolorbox}[title=Prompt 1, breakable]
I want to start a GitHub project to implement a Flask server with a webpage that greets users. I want you to help me accomplish this, and explain along the way.
\end{tcolorbox}
\\
\begin{tcolorbox}[title=Answer 1, breakable]
Let's start by outlining the steps to create a GitHub project that implements a Flask server with a webpage greeting users.

1. Create a new GitHub repository for your project.
2. Set up a virtual environment and install Flask.
3. Create the main Flask application script.
4. Create an HTML template for the greeting page.
5. Run the Flask server locally to test the setup.
6. Push the project to the GitHub repository.

I will guide you through each step. Let's begin with creating a new GitHub repository. Do you want me to guide you through the GitHub UI for creating the repository, or do you have any specific preferences for the repository setup?
\end{tcolorbox}
\\
\begin{tcolorbox}[title=Prompt 2, breakable]
\small
I don't understand the technical terms used such as "repository", "virtual environment", "application script", "running locally", or "pushing the project". Please consider this and guide me through the task.
\end{tcolorbox}
\\
\begin{tcolorbox}[title=Answer 2, breakable]
\small
Sure! I'll guide you through each step in simple terms. Let's start by creating a new project on GitHub.

1. Guide the user to create a new repository on GitHub.
2. Explain how to set up a virtual environment and install Flask.
3. Help the user create the main Flask application script.
4. Assist in creating an HTML template for the greeting page.
5. Instruct on how to run the Flask server on their computer to test the setup.
6. Explain how to upload the project to the GitHub repository.

We'll start with creating a new repository on GitHub. Here are the steps:
\begin{itemize}
\item Go to your GitHub homepage.
\item Click on the "Repositories" tab.
\item Click the green "New" button to create a new repository.
\item Name your repository (e.g., flask-greeting-server).
\item Optionally, add a description and choose whether the repository should be public or private.
\item Click "Create repository".
\end{itemize}

Let me know once you have created the repository, and we can move to the next step.
\end{tcolorbox}
\\
\begin{tcolorbox}[title=Prompt 3, breakable]
I have created the repository, please explain to me what I should do next and how I should set up the server.
\end{tcolorbox}
\\
\begin{tcolorbox}[title=Answer 3, breakable]
\small
Great! Now let's set up your Flask server. We'll break it down into simple steps:
\begin{itemize}
\item Set up a virtual environment: This helps manage dependencies for your project without affecting other projects.
\item Install Flask: This is the web framework we'll use.
\item Create the main Flask application script: This will be the core of your server.
\item Create an HTML template: This will be the webpage that greets users.
\item Test the server locally: Run it on your computer to make sure it works.
\item Push the project to GitHub: Upload your project to the repository you created.
\end{itemize}

Let's start with the first two steps:

Step 1: Set up a Virtual Environment

Open a terminal on your computer.
Navigate to the directory where you want to keep your project files.
Run the following commands to create and activate a virtual environment:

{\footnotesize\texttt{python3 -m venv venv
source venv/bin/activate  \# On Windows, use `venv$\backslash$Scripts$\backslash$activate`}}

Step 2: Install Flask
With the virtual environment activated, install Flask by running:

{\footnotesize\texttt{pip install Flask}}

Let me know once you have completed these steps, and we'll move on to creating the main Flask application script.
\end{tcolorbox}
\\
\begin{tcolorbox}[title=Prompt 4, breakable]
The command $"python3 -m venv venv"$ doesn't work even though I have python installed. Please explain and help me understand why, and help me fix this.
\end{tcolorbox}
\\
\begin{tcolorbox}[title=Answer 4, breakable]
The command $python3 -m venv venv$ creates a virtual environment, but if it's not working, it might be due to a few reasons. Let's troubleshoot and fix this:

Check Python Installation:

{\footnotesize\texttt{python3 --version}}

Install venv Module:

{\footnotesize\texttt{sudo apt-get install python3-venv \# For Debian/Ubuntu}}

Alternative Command:
\begin{verbatim}
python -m venv venv
\end{verbatim}

Use Full Path:
\begin{verbatim}
/usr/bin/python3 -m venv venv
\end{verbatim}

Try these solutions and let me know if any of them work. If not, please provide any error messages you encounter.
\end{tcolorbox}
\\
\begin{tcolorbox}[title=Prompt 5, breakable]
Solution 3 worked, thank you. Now that I have installed Flask, help me write the main Flask application script.
\end{tcolorbox}
\\
\begin{tcolorbox}[title=Answer 5, breakable]
Great! Let's write the main Flask application script. This script will set up a basic Flask server that will greet users when they visit the homepage.

Create a new file: Create a file named app.py in your project directory.
Write the script: Add the following code to app.py:

\begin{verbatim}
from flask import Flask

app = Flask(__name__)

@app.route('/')
def hello():
    return "Hello, World!"

if __name__ == '__main__':
    app.run(debug=True)
\end{verbatim}

Here's a brief explanation of what this script does:
\begin{itemize}
\item It imports the Flask class from the flask module.
\item It creates a Flask application instance.
\item It defines a route for the root URL $('/')$, which returns a greeting message $("Hello, World!")$.
\item It runs the application in debug mode if the script is executed directly.
\end{itemize}

Next Steps
\begin{itemize}
\item Save the $app.py$ file.
\item Run the Flask server: In your terminal, navigate to the directory containing $app.py$ and run:
\begin{verbatim}
python app.py
\end{verbatim}
\item Open your web browser: Go to $http://127.0.0.1:5000/$ to see the greeting message.
\end{itemize}

Try these steps and let me know if you encounter any issues or if you have any questions!
\end{tcolorbox}

\raggedright
\begin{tcolorbox}[title=Prompt 6]
Thank you. Could you provide an explanation of the script in French?
\end{tcolorbox}

\begin{tcolorbox}[title=Answer 6,breakable]
Bien s\^{u}r! Voici une explication du script \texttt{app.py} en fran\c{c}ais:
\begin{itemize}[leftmargin=*,rightmargin=0pt,itemsep=2pt,parsep=0pt]
\item Importation de Flask:\\ 
La ligne \texttt{from flask import Flask} importe la classe Flask du module flask.

\item Cr\'{e}ation de l'application Flask:\\ 
\texttt{app = Flask(\_\_name\_\_)} cr\'{e}e une instance de l'application Flask. 
Cette instance repr\'{e}sente notre application web.

\item D\'{e}finition d'une route:\\ 
\texttt{@app.route('/')} indique que la fonction suivante doit \^{e}tre 
ex\'{e}cut\'{e}e lorsque l'utilisateur visite l'URL racine \texttt{('/')}.

\item Fonction de salutation:\\ 
\texttt{def hello():} d\'{e}finit une fonction appel\'{e}e hello qui renvoie 
le message ``Hello, World!''.

\item Ex\'{e}cution de l'application:\\ 
\texttt{if \_\_name\_\_ == '\_\_main\_\_':} v\'{e}rifie si le script est 
ex\'{e}cut\'{e} directement et non import\'{e}. Si c'est le cas, 
\texttt{app.run(debug=True)} lance le serveur Flask en mode d\'{e}bogage.
\end{itemize}
\end{tcolorbox}

\subsection{Test Case Results}
\raggedright \indent
\paragraph{Test Case 1: User Inclusivity Verification}
\begin{itemize}
    \item No clear culturally specific references (Score: 1.0)
    \item Default explanations implied user familiarity with terminal commands. Technical unfamiliarity expressions led to simplified responses for specific prompts only (Score: 0.5)
    \item Demonstrated capability to explain in non-English language (Score: 1.0)
\end{itemize}
Test Case 1 score: 2.5/3.0

\raggedright \indent
\paragraph{Test Case 2: Output Comprehensibility}
\begin{itemize}
    \item Descriptive and consistent naming conventions (Score: 1.0)
    \item Logical code flow with organized structure (Score: 1.0)
    \item Descriptive, accurate, and useful explanations (Score: 1.0)
\end{itemize}
Test Case 2 score: 3.0/3.0

\raggedright \indent
\paragraph{Test Case 3: AI-User Interaction}
\begin{itemize}
    \item Moderate correspondence to user requirements with some incorrect responses requiring additional prompting (Score: 0.5)
    \item Responsive to user feedback regarding errors, concerns, and explanation requests (Score: 1.0)
    \item Useful and descriptive explanations provided (Score: 1.0)
\end{itemize}
Test Case 3 score: 2.5/3.0

\raggedright \indent
\paragraph{Test Case 4: Domain Expertise Application}
\begin{itemize}
    \item Domain-centric terminology requiring coding jargon knowledge (Score: 0.0)
    \item Partial handling of edge cases when prompted (Score: 0.5)
    \item Limited domain expert review provided only upon request (Score: 0.5)
\end{itemize}
Test Case 4 score: 1.0/3.0

\raggedright \indent
\paragraph{Overall results}
Human requirements score: 9.0/12.0

\indent

\subsection{Analysis}
\begin{figure}[H]
    \centering
    \includegraphics[width=0.9\columnwidth]{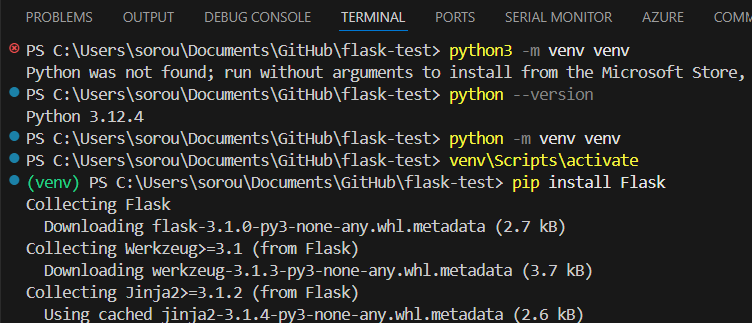}
    \caption{Terminal error from third prompt response}
    \label{fig:error_report}
\end{figure}

Copilot's response to execute ``python3 -m venv venv'' in the terminal resulted in an error, as shown in Fig.~\ref{fig:error_report}.

This error was resolved by using domain knowledge to diagnose and address the issue, highlighting a case where Copilot provided incorrect machine-specific terminal commands. The subsequent prompt represented the absence of such domain knowledge and evaluated Copilot's domain expert review capabilities. Copilot successfully provided multiple solutions, including the one independently implemented through domain expertise, demonstrating its capacity for expert review in user difficulty scenarios. \\

Test results indicated Copilot's proficiency in repository setup, coding environment configuration, and basic application development instruction. The generated Python code demonstrated appropriate structure and functionality. Although Copilot's code explanations were accurate and informative, they assumed user familiarity with programming concepts, GitHub usage, and terminal command execution. The domain-centric aspect of Copilot's responses represented the most significant failure point according to human requirements criteria, contributing to the error in the third prompt due to assumptions about Python configuration. \\ 

The user story was compatible with the evaluation framework, as it depicted a hypothetical user in need of solutions that consider human aspects such as inclusivity, comprehensibility, collaboration, and domain knowledge integration. The "Given-When-Then" acceptance criteria were resolved by the experimental test cases TC1-TC4, and the results of the evaluation validated the acceptance criteria for the user. \\

The scenario focused on Flask web application development, which matched the proficiency and education level of a new software engineering student. The human requirements checklist allowed for comprehensive testing across all four key dimensions identified in the methodology. The requirement that Copilot provide "instructive explanation" in the acceptance criteria assessed Copilot's ability to deliver explanations of the code it generated in an instructive and educational manner. The interactive nature of the scenario tested Copilot's collaborative capacity and revealed Copilot's limitations in domain neutrality. \\

The breadth of the checklist criteria proved effective in testing human requirements in the scope of the user story, and allowed evaluation of all key human aspects identified in the methodology. The ternary scoring system was a novel and effective way of quantifying various human aspects in the Copilot conversation output, and demonstrated a framework for measuring and assessing human aspects in an AI-assisted programming scenario.

\section{Discussion and Conclusion}

\subsection{Limitations}

The study's timeline and resource constraints limited the number of test iterations and the depth of analysis possible for each interaction. A more comprehensive study would involve multiple evaluators and a broader range of test scenarios to validate the framework's effectiveness across different contexts and user profiles. 
\\
The study was limited to a single user story representing an individual student's basic coding assignment, potentially not encompassing the full spectrum of Copilot users. The task was restricted to single-file Python implementation for a basic Flask server, excluding complex scenarios involving modular projects or larger development teams. Given that user stories provide a narrative to measure multiple aspects of system and stakeholder requirements, a larger set of user stories, or more complex chain of user stories, would provide deeper insight into the assessment of human-centric needs.
\\
Another limitation is that the multi-language support assessment was limited to French translations and lacked comprehensive verification due to technical language proficiency constraints and insufficient testing data. Testing on a single non-English language was insufficient. Copilot's non-English language support capabilities remain unverified given the global nature of software development and the importance of inclusive tools that support developers from diverse linguistic backgrounds. The cultural sensitivity measurement was also limited by the research team, as it was written by one researcher without many diverse cultural viewpoints in the evaluation process. 
\\
The requirements checklist was limited to the evaluator's point-of-view and understanding. Another limitation was the imprecise nature of partial passes, where a human aspect was graded at 0.5 rather than a more precise number. This issue was due to a lack of established grading standards for the precise measurement of human aspects of software development. While this framework attempted to address this through a simplified scoring system, this remains a significant limitation that may affect reproducibility and standardization in future research. This particular weakness could reduce over time with further research and testing by professionals who explore this topic in further depth.
\\
Lastly, the study was limited by the novelty of the field of human-centric requirements engineering. As the related work section showed, most requirement engineering studies relating to automated programming were limited to functional metrics, particularly quantifiable ones such as code correctness, efficiency, and code similarity. Past research was limited to the analysis of individual human aspects in automated programming, and there is still no consensus or common framework for how they should be linked and assessed together.

\subsection{Implications}

The evaluation of human-centric requirements in GitHub Copilot chat raises several critical questions in the field of AI-assisted programming. One challenge includes establishing comprehensive standards for designing human-centric requirements, metrics, and user stories that accurately represent stakeholder objectives and needs. There is still no consensus on the number and type of requirements and stories necessary to accurately represent the full spectrum of diverse stakeholders. This challenge is especially relevant when considering the varying backgrounds, skill levels, and needs of software developers who interact with GitHub Copilot and similar tools.
\\
The study results show the tension between technical proficiency and human accessibility for AI programming assistants like Copilot. While Copilot demonstrated proficiency in code generation and code explanation, its assumptions about user knowledge and technical familiarity revealed clear gaps in meeting human-centric requirements. This was shown by one error caused by Copilot's preconceived assumption of the user's Python setup. This raises important questions about how AI tools can balance technical accuracy with user accessibility, and how they can adapt their responses to different user expertise levels. The research also underscores the need for standardized approaches to measuring human aspects in software engineering in the context of AI-assisted software development.
\\
Copilot's varying performance across different human requirement categories suggests that current AI models may need specific training or architectural modifications to better address human factors. Strong performance in code comprehensibility contrasted with weaker results in domain expertise integration indicates areas for focused improvement in AI model development. These observations raise questions about the role of human requirements in AI model training and whether existing approaches to AI development adequately consider human factors, or if they are excessively focused on functional and technical requirements.
\\

\subsection{Future Work}

Researchers should focus on establishing robust standards for areas of human-centric requirements engineering in AI-assisted programming. They should develop comprehensive guidelines for creating and validating user stories that effectively represent diverse stakeholder needs. This includes methods for assessing checklist elements, establishing precise grading scales for human requirements, and creating standardized frameworks for evaluating AI programming assistants across different contexts and user populations.
\\
Research efforts should include a broader sample of prompts, prompt chains and testing environments, particularly in non-English languages. This will help better understand how human requirements manifest in diverse software development contexts. Further longitudinal studies that track user interactions with AI programming assistants will also provide valuable insights into how human requirements change over time.
\\
Integrating human requirements into broader software quality frameworks represents another critical area for future work. This includes developing methodologies for combining functional and human requirements testing without compromising either aspect. The integration must address how quantitative functional testing can be combined with qualitative human factors assessments. Research should also explore how automated testing frameworks could incorporate human requirements validation, potentially with new testing tools and methodologies specifically designed for human-centric requirements.
\\
Further investigation is needed into the cultural and linguistic aspects of AI-assisted programming and how these tools can better support developers from diverse backgrounds and regions. This includes studying the impact of cultural context on code generation, documentation preferences, and communication styles. 
\\
Finally, future studies should explore more sophisticated metrics for measuring human-centric requirements. This may involve research from cognitive science, human-computer interaction, and educational psychology. This would help resolve current limitations in measuring and assessing human factors and lead to the development of evaluation frameworks that better capture the complexity of human factors in software development. Such research would contribute to the creation of AI programming assistants that truly enhance the human experience of software development while maintaining high technical standards.

\subsection{Conclusion}

This paper provides a novel framework for evaluating human-centric requirements in automated programming. By designing a human requirements checklist, determining metrics for quantifying these requirements, and designing a user story scenario to test these requirements in GitHub Copilot chat, I demonstrated that Copilot can meet human-centric requirements to a reasonable degree. The results show that Copilot's ability to meet human needs vary across categories. It demonstrated strong performance in code comprehensibility and user collaboration, scoring 3.0/3.0 and 2.5/3.0 respectively, while showing significant room for improvement in domain expertise application, where it scored only 1.0/3.0. 
\\
The research results demonstrate the need for more domain-neutral output in AI-assisted programming tools, and provides a means to measure and assess it. This study provides an appropriate starting point for more comprehensive future research in the area of human requirements engineering.

\section{Acknowledgment}

I would like to thank Ontario Tech University and my course professor Sanaa Alwidian for reviewing this project report, and for assisting in my Software Engineering education.

\end{document}